\documentclass[onecollarge,natbib]{svjour2}
\bibpunct{[}{]}{;}{n}{}{,} % to get "[numbered]" references from natbib
\smartqed  % flush right qed marks, e.g. at end of proof
\usepackage{graphicx}
\usepackage{amsmath}
\journalname{Few-Body Systems (EFB22)}
\begin{document}

\title{Theoretical description of deeply virtual Compton scattering off $^3$He
}

%%%%%%%%%%%%%%%%%%%%%%%%%%%%%%%%%%%%%%%%%5%

\author{M. Rinaldi         \and
        S. Scopetta %etc.
}


\institute{M. Rinaldi \at
	Dipartimento di Fisica, Universit\`{a} degli studi di Perugia 
and INFN sezione di Perugia, Via A. Pascoli 06100 Perugia, Italy
 \\
%Tel.: +39 075 585 2793\\
\email{matteo.rinaldi@pg.infn.it}           %  \\
\and
S. Scopetta \at
Dipartimento di Fisica, Universit\`{a} degli studi di Perugia 
and INFN sezione di Perugia, Via A. Pascoli 06100 Perugia, Italy
\\
%Tel.: +39 075 585 2721
\email{sergio.scopetta@pg.infn.it} 
}

\date{Received: date / Accepted: date}

\maketitle

\begin{abstract}
Recently, coherent deeply virtual
Compton scattering (DVCS) off
$^3$He nuclei has been proposed
to access the neutron generalized parton distributions (GPDs).
In Impulse Approximation (IA) studies,
it has been shown, in particular, that the sum of 
the two leading twist, quark helicity conserving
GPDs of $^3$He, $H$ and $E$, at low momentum transfer, is 
dominated by the neutron contribution, so that
$^3$He is very promising for the extraction of the neutron information. 
Nevertheless, such an extraction could be not trivial.
A technique, able to take into account the nuclear effects 
included in the IA analysis in the extraction procedure, has been 
therefore developed.
In this work, the IA calculation of the spin dependent
GPD $\tilde H$ of $^3$He is presented for the first time. This quantity
is found to be largely dominated, at low momentum
transfer, by the neutron contribution, which
could be extracted using arguments similar to the ones previously proposed
for the other GPDs.
The known forward limit of the IA calculation of $\tilde H$,
yielding the polarized parton distributions of $^3$He, is correctly
recovered.
The knowledge of the GPDs $H, E$ and $\tilde H$ of $^3$He will allow now
the evaluation of the cross section asymmetries which
are relevant for coherent DVCS off $^3$He 
at Jefferson Lab kinematics, an important step
towards the planning of possible experiments.

\end{abstract}
\keywords{Three body systems \and Generalized parton distributions}
%\PACS{ 13.60.Hb \and  21.45-v \and 14.20.Dh}
\vskip2mm
%\newpage

%\section{}
%\label{intro}
Initially introduced in Ref. \cite{uno},
Generalized Parton Distributions (GPDs),
among other features,
represent a crucial tool to shed light on 
the so called ``Spin Crisis'' problem.
As a matter of fact, GPDs measurements will
allow to access the parton total angular momentum \cite{due}.  
By subtracting from the latter the helicity quark contribution, 
measured in other
hard processes, the parton orbital angular momentum (OAM)
could be then estimated. 

\begin{figure*}[t]
\vspace{6.5cm}
%\special{psfile=ff_fermi.eps hoffset=0
%voffset=70
%hscale= 45 vscale= 45 
%angle=0}
\includegraphics{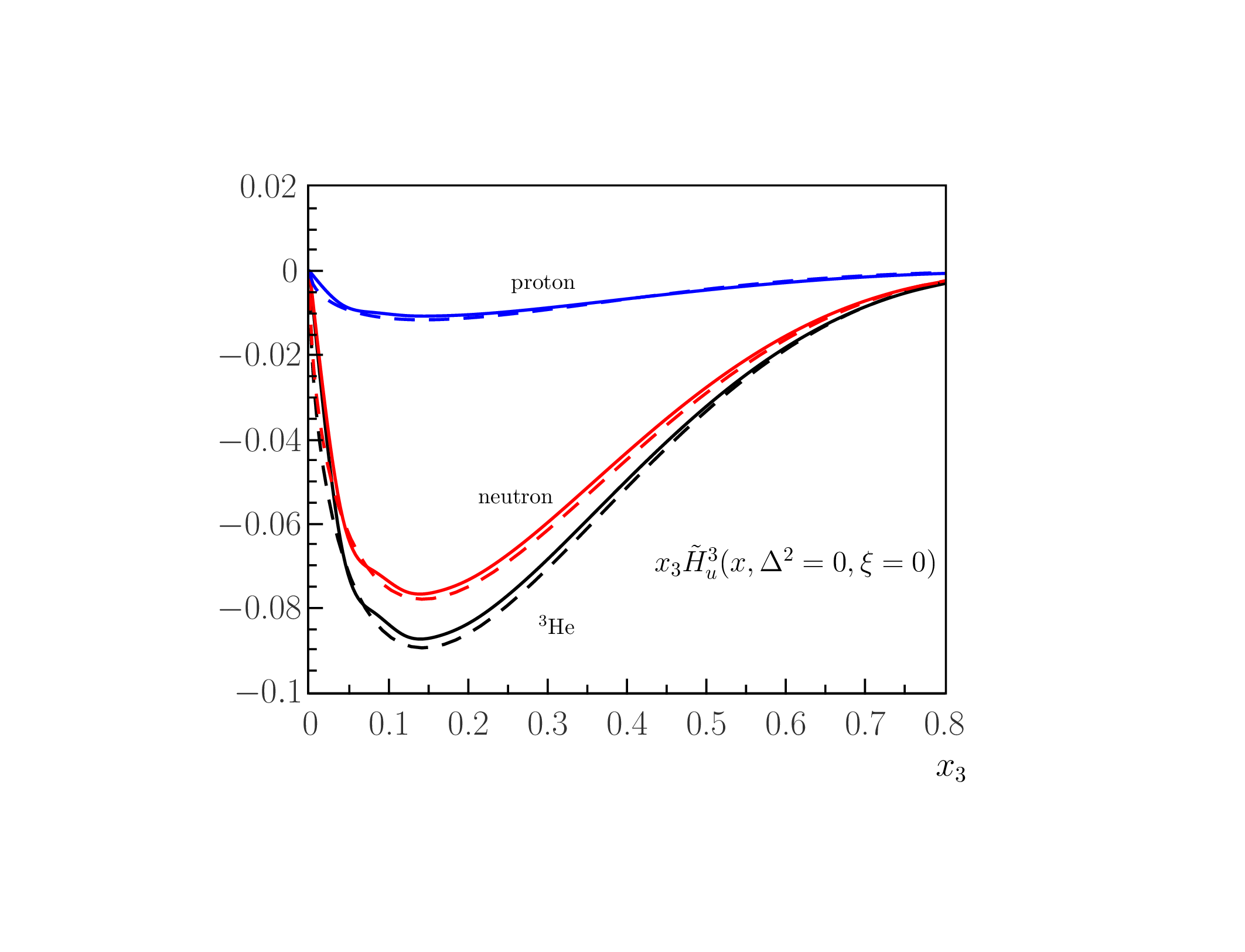}
\includegraphics{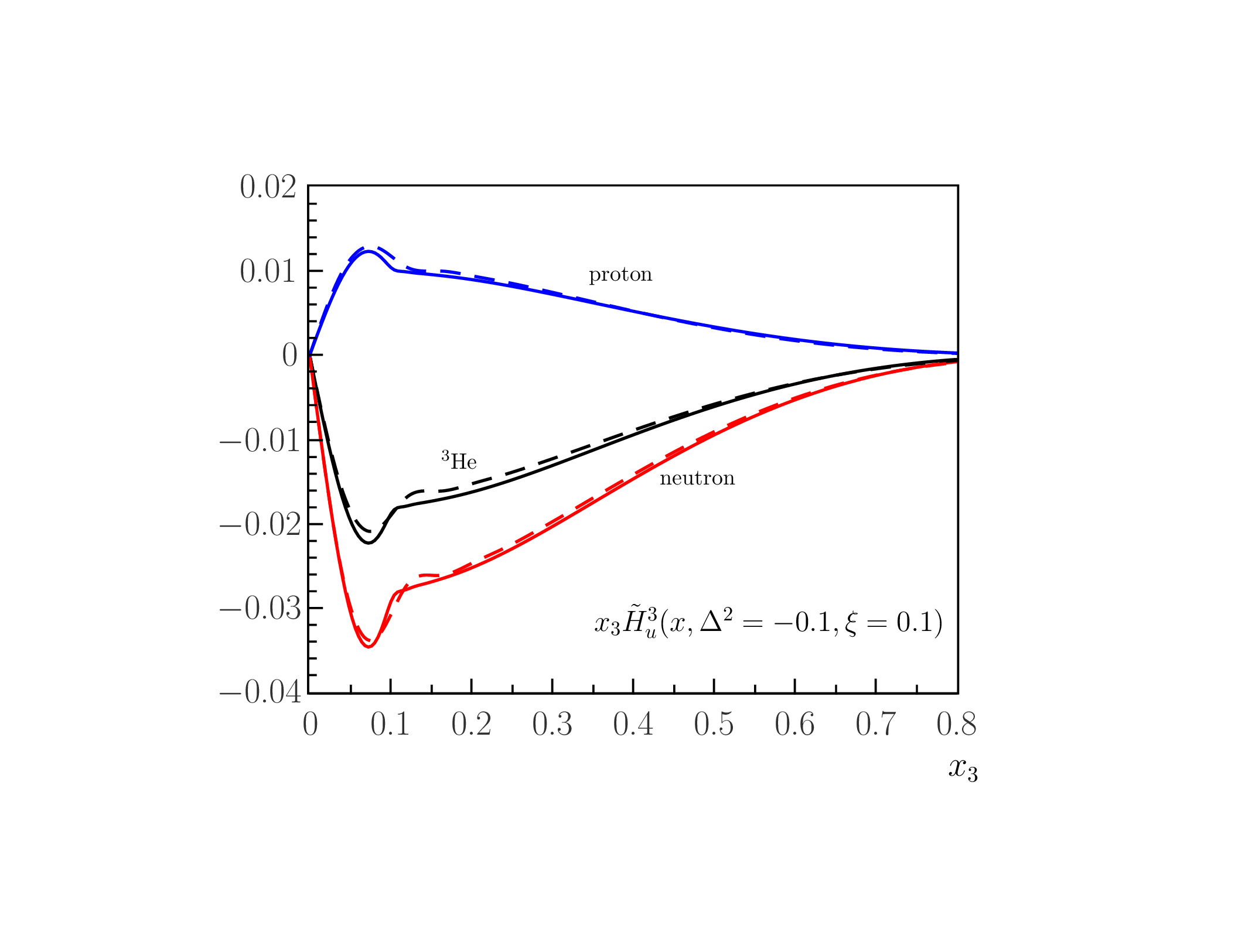}
%\end{figure*}
%
%\begin{figure*}[t]
%\vspace{7.5cm}
%\vskip-3.1cm
%\vskip-9.3cm
\caption{ 
The $q=u$ flavor GPD $x_3 \tilde{H}^{3,u}(x,\Delta^2,\xi)$, 
where $x_3 = (M_3/M) x$
and $\xi_3 = (M_3/M) \xi$, shown in the forward limit (left panel)
and at $\Delta^2 = -0.1
\ \mbox{GeV}^2$ and $\xi_3=0.1$ (right panel), 
together with the neutron and the proton
contribution. Solid lines represent the full IA result, Eq. (\ref{new}),
while the dashed ones correspond to the approximation
Eq. (\ref{sgu}).}
\end{figure*}

Deeply Virtual Compton Scattering 
(DVCS), i.e. the reaction
$eH \longmapsto e'H' \gamma$ \ when \ $Q^2\gg M^2$ ($Q^2=-q \cdot q$ \ 
is the momentum transfer
between the leptons $e$ and 
$e'$, $\Delta^2$ the one between hadrons $H$ and $H'$ with
momenta $P$ and $P'$, and
$M$ is the nucleon mass), is considered
the cleanest process to access GPDs.
Another relevant kinematical variable is the so called
skewedness, $\xi = - \Delta^+/(P^+ + P^{'+})$ 
\footnote{In this paper, $a^{\pm}=(a^0 \pm a^3)/\sqrt{2}$.}.
%%%%%% NUCLEI
Despite severe difficulties to extract GPDs from  experiments, 
data for proton and nuclear targets are being analyzed,
see, i.e., Refs. \cite{data1,data2}.
The measurement of GPDs for nuclei
could be crucial to distinguish
between different models of nuclear
medium modifications of the nucleon structure,
an impossible task in the analysis of DIS experiments only
(this discussion started in Ref. \cite{deu}).
As always, the neutron measurement, which requires
nuclear targets, is important because it permits, 
together with the proton one,
a flavor decomposition of GPDs.
In studying observables related to 
the neutron polarization, $^3$He plays a special role, 
due to its spin structure (see, e.g., Refs. \cite{3He,old}). 
This is true in particular for GPDs.
Indeed, among the light nuclei
$^3$He is the only one for which the quantity
$\tilde{G}_M^{3,q}(x,\Delta^2,\xi) = H^{3}_q(x,\Delta^2,\xi)+
E^{3}_q(x,\Delta^2,\xi)$, i.e., the sum of its GPDs $H_q$ and
$E_q$, could be dominated by the neutron, being 
the isoscalar targets
$^2$H and $^4$He not suitable to this aim, as it has been discussed
in Ref. \cite{noi}.
In the same papers it has been also shown 
to what extent this fact can be used to extract the neutron information.

The formal treatment of $^3$He GPDs 
in Impulse Approximation (IA) can be found in Refs. \cite{scopetta},
where, for the GPD $H$ of $^3$He,
$H_q^3$, 
a convolution-like equation
in terms of the corresponding
nucleon quantity has been obtained.
The treatment has been later extended to
$\tilde{G}_M^{3,q}$
(see Ref. \cite{noi} for details),
and to $\tilde{H}^3_q$,
yielding
%\newpage
\begin{eqnarray}
 \tilde G_M^{3,q}(x,\Delta^2,\xi)  = 
\sum_N
\int dE 
\int d\vec{p}~
{\left [ P^N_{+-,+-}
%(\vec p,\vec p\,',E)  
-
P^N_{+-,-+} \right](\vec p,\vec p\,',E) }
{\xi' \over \xi}
\tilde G_M^{N,q}(x',\Delta^2,\xi'),
\end{eqnarray}
\vskip-3mm
\noindent
and
\begin{eqnarray}
{\tilde H_{q}^3(x,\Delta^2,\xi)}  = 
\sum_N
\int dE 
\int d\vec{p}
\,
{\left [ P^N_{++,++}
%(\vec p,\vec p\,',E)  
-
P^N_{++,--} \right](\vec p,\vec p\,',E) }
{\xi' \over \xi}
{\tilde H^{N}_q(x',\Delta^2,\xi')}~,
\label{new}
\end{eqnarray}
\vskip-3mm
\noindent
respectively.

In the last two equations, $x'$ and $\xi'$ are the variables 
for the bound nucleon 
GPDs, $p \, (p'= p + \Delta)$
is its 4-momentum in the initial (final) state and, eventually, 
proper components appear of the spin dependent,
one body off diagonal spectral function:

\begin{eqnarray}
 \label{spectral1}
 P^N_{SS',ss'}(\vec{p},\vec{p}\,',E) 
= 
\dfrac{1}{(2 \pi)^6} 
\dfrac{M\sqrt{ME}}{2} 
\int d\Omega _t
%\\
% \times   
\sum_{\substack{s_t}} \langle\vec{P'}S' | 
\vec{p}\,' s',\vec{t}s_t\rangle_N
\langle \vec{p}s,\vec{t}s_t|\vec{P}S\rangle_N~,
%\nonumber 
\end{eqnarray}
where $S,S'(s,s')$ are the nuclear (nucleon) spin projections
in the initial (final) state, respectively,
and $E= E_{min} +E_R^*$, 
being $E^*_R$ the excitation energy 
of the full interacting two-body recoiling system.
The main quantity appearing in the definition
Eq. (\ref{spectral1}) is
the intrinsic overlap integral
\begin{equation}
\langle \vec{p} ~s,\vec{t} ~s_t|\vec{P}S\rangle_N
=
\int d \vec{y} \, e^{i \vec{p} \cdot \vec{y}}
\langle \chi^{s}_N,
\Psi_t^{s_t}(\vec{x}) | \Psi_3^S(\vec{x}, \vec{y})
 \rangle~
\label{trueover}
\end{equation}
between the wave function
of $^3$He,
$\Psi_3^S$,  
and the final state, described by two wave functions: 
{\sl i)}
the
eigenfunction $\Psi_t^{s_t}$, with eigenvalue
$E = E_{min}+E_R^*$, of the state $s_t$ of the intrinsic
Hamiltonian pertaining to the system of two {\sl interacting}
nucleons with relative momentum $\vec{t}$, 
which can be either
a bound 
%(deuteron) 
or a scattering state, and 
{\sl ii)}
the plane wave representing 
the nucleon $N$ in IA.
For a numerical evaluation of Eqs. (1) and (2),
the overlaps, Eq. (4), appearing in Eq. (3)
and corresponding to the analysis presented in Ref.
\cite{overlap} in terms of 
AV18  \cite{pot} wave functions
\cite{AV18}, 
have been used. For the nucleonic GPDs, a simple
model for $\tilde G_M^{N,q}$ 
and $\tilde H^{N}_q$
\cite{Rad1}, properly extended to evaluate
also spin dependent GPDs
(see Ref. \cite{noi} for details), has been used.
%
%\vskip-7.01mm
It is worth noticing that Eq. (\ref{new}) and the results
of its numerical evaluation are presented here for 
the first time.
Since there are no $^3$He data available, 
it is possible to verify only a few general 
GPDs properties, i.e., the forward limit and the first moments.
In particular the calculation of $H^{3}_q(x,\Delta^2,\xi)$ 
fulfills these constraints
\cite{scopetta}. 
In the  $\tilde G_M^{3,q}(x,\Delta^2,\xi)$ case, since there is 
no observable forward limit
for $E^{3}_q(x,\Delta^2,\xi)$, the only possible check is the first moment:
$
\sum_q \int dx \, \tilde G_M^{3,q}(x,\Delta^2,\xi) = G_M^3(\Delta^2);
$
where $G_M^3(\Delta^2)$ is the magnetic form factor (ff) 
of $^3$He.
The result obtained has been found to be in agreement with previous
calculations (e.g. the one-body part of the
AV18 calculation
presented in Ref. \cite{schiavilla}) and, 
for the values of $\Delta^2$ which are relevant for the 
coherent process under investigation here,
i.e., $-\Delta^2 \le 0.15$ GeV$^2$,
our results compare well also with the data.

Let us discuss now our calculation of $\tilde H^3_q$, presented here
for the first time. First of all,
we checked that the forward limit of our expression,
Eq. (\ref{new}), reproduces formally and numerically
the formalism obtained in Ref. \cite{old} for polarized DIS
off $^3$He. On the other hand,
the first moment of $\tilde H^3_q$ is related
to the axial form factor of $^3$He, an observable
poorly known which does not permit therefore a consistency check.
With the comfort of the fulfillment of the forward limit constraint
we can now proceed to analyze the proton and neutron
contributions to the $^3$He observable. 
Since $\tilde H_q^3$ is measured using a polarized target, it should
be dominated by the neutron contribution.
Let us show now to what extent
this feature is obtained and how, thanks to this observation,
the neutron information can be extracted.

The results of the numerical evaluation of Eq. (\ref{new}) are
presented in Fig. 1. 
In the forward limit, 
the neutron contribution strongly dominates the
$^3$He quantity, but 
increasing $\Delta^2$ 
the proton contribution grows up
(see Fig. 1, solid lines in both panels), 
in particular for the $u$ flavor.
It is therefore necessary to introduce
a procedure to safely extract the neutron information from 
$^3$He
data. This can be done by observing that
Eq. (\ref{new}) can be written as
\begin{eqnarray}
\tilde H^{3}_q(x,\Delta^2,\xi) =   
\sum_N \int_{x_3}^{M_A \over M} { dz \over z}
h_N^3(z, \Delta^2, \xi ) 
\tilde H^{N}_q \left( {x \over z},
\Delta^2,
{\xi \over z},
\right)~,
%\nonumber
\end{eqnarray}
%\vskip-.9cm
where $h_N^3(z, \Delta^2, \xi )$ 
is a ``light cone spin dependent off-forward momentum
distribution'' which
is strongly peaked around $z=1$
close to the forward limit. 
Therefore, in this region,
for $x_3 = (M_A/M) x \leq 0.7$ one has:

\begin{eqnarray}
\tilde H^{3}_q(x,\Delta^2,\xi) 
& \simeq &   
\sum_N 
\tilde H^{N}_q \left( x, \Delta^2, {\xi } \right)
\int_0^{M_A \over M} { dz }
h_N^3(z, \Delta^2, \xi ) 
\nonumber
\\
& = &
G^{3,p,point}_A(\Delta^2) 
\tilde H^{p}_q(x, \Delta^2,\xi) 
+ 
G^{3,n,point}_A(\Delta^2) 
\tilde H^{n}_q(x,\Delta^2,\xi)~. 
%\nonumber
\label{sgu}
\end{eqnarray}
\vskip-1mm
Here, the axial point like ff, 
$G^{3,N,point}_A(\Delta^2)=\int_0^{M_A \over M} dz \, h_N^3(z,\Delta^2,\xi), $ 
which would give the nuclear axial ff if the proton and the neutron were 
point-like
particles, are introduced.
These quantities, at small values of $\Delta^2$, depend 
weakly on the potential used in the calculation, 
so that the theoretical error in their evaluation is small.
This can be realized observing that, in the forward limit, they reproduce
the so called ``effective polarizations'' of the protons ($p_p$) 
and the neutron ($p_n$) in $^3$He, whose values are rather similar 
if evaluated within different nucleon nucleon potentials 
(see Refs. \cite{3He,old,overlap} for a comprehensive discussion). 
In particular, within the AV18 potential under scrutiny here,
the values $p_n=0.878$ and $p_p= -0.023$ are obtained.
Eq. (\ref{sgu}) can now be used to extract the
neutron contribution from possible sets of data
for the proton and for $^3$He:
\vskip-5mm
\begin{eqnarray}
\label{extr}
\tilde H^{n,extr}(x, \Delta^2,\xi)  \simeq  
{1 \over G^{3,n,point}_A(\Delta^2)} 
\left\{ \tilde H^3(x, \Delta^2,\xi) 
 -  
G^{3,p,point}_{A}(\Delta^2) 
\tilde H^p(x, \Delta^2,\xi) \right\}~.
\end{eqnarray}

The comparison we have done between the free neutron GPDs, used 
as input in the calculation,
and  the ones extracted using our calculation for
$\tilde H^3$ and the proton model for $\tilde H^p$, shows that the
procedure works nicely even beyond the forward limit. The only theoretical
ingredients are the axial point like ffs, 
which, as explained above, are under good theoretical control.
The procedure works for $x \leq 0.7$,
where possible data are expected from JLab.

In closing, we have shown that coherent DVCS off $^3$He at
low momentum transfer $\Delta^2$ is an ideal process to access 
the neutron GPDs;
if data were taken at higher $\Delta^2$, a relativistic treatment 
\cite{ema}
and/or the inclusion of many body currents, beyond the present IA scheme, 
should be implemented. 
The next step of this investigation will be the evaluation of cross section
asymmetries relevant to DVCS experiments at JLab kinematics, using
the obtained theoretical GPDs $H, E$ and $\tilde H$ of $^3$He.
At the beginning, the leading twist analysis of DVCS for a spin 1/2 target,
presented in Ref. \cite{diet}, will be performed.  

This work was supported in part by the Research Infrastructure
Integrating Activity Study of Strongly Interacting Matter (acronym
HadronPhysic3, Grant Agreement n. 283286) under the Seventh Framework
Programme of the European Community.


\begin{thebibliography}{99}

%\cite{Mueller:1998fv}
\bibitem{uno}
Mueller,~D. {\sl et al.}:
%, D.~Robaschik, B.~Geyer, F.~M.~Dittes and J.~Horejsi,
  %``
Wave functions, evolution equations and evolution kernels from light ray operators of QCD.
%''
Fortsch.\ Phys.\  {\bf 42}, 101 (1994);
%  [hep-ph/9812448];
  %%CITATION = HEP-PH/9812448;%% 
\bibitem{due}
  %\cite{Radyushkin:1996nd}  
Radyushkin,~A.~V.
 %``
Scaling limit of deeply virtual Compton scattering.
%''
Phys.\ Lett.\ B {\bf 380}, 417 (1996);
%  [hep-ph/9604317];
  %%CITATION = HEP-PH/9604317;%%
    %\cite{Ji:1996ek}
Ji,X.~-D.
  %``
Gauge invariant decomposition of nucleon spin and its spin - off
.
%''
Phys.\ Rev.\ Lett.\  {\bf 78}, 610 (1997).
%  [hep-ph/9603249].
  %%CITATION = HEP-PH/9603249;%%
  
%\bibitem{rassegne}
%%\cite{Diehl:2003ny}
%Diehl,~M. Phys. Rept. 388, 41 (2003);
% %``Generalized parton distributions,''
%  [hep-ph/0307382].
%  %%CITATION = HEP-PH/0307382;%%
%  %\cite{Belitsky:2005qn}  
%Belitsky,~A.~V. and Radyushkin,~A.~V., Phys. Rept. 418, 1 (2005); 
%%[hep-ph/0504030].
%  %%CITATION = HEP-PH/0504030;%%
%%\cite{Boffi:2007yc}  
%Boffi,~S. and Pasquini,~B. Riv. Nuovo Cim. 30, 387 (2007). 
% %``Generalized parton distributions and the structure of the nucleon,''
%%  [arXiv:0711.2625 [hep-ph]].
%  %%CITATION = ARXIV:0711.2625;%%

\bibitem{data1}
%\cite{Airapetian:2009bm}
%Airapetian,~A. {\it et al.}  [HERMES Collaboration],
%  %``Measurement of azimuthal asymmetries associated with deeply virtual Compton scattering on an unpolarized deuterium target,''
%  Nucl.\ Phys.\ B {\bf 829} (2010) 1;
%%  [arXiv:0911.0095 [hep-ex]].
%  %%CITATION = ARXIV:0911.0095;%%
%%\cite{Airapetian:2009bi}
%  %``Nuclear-mass dependence of azimuthal beam-helicity and beam-charge asymmetries in deeply virtual Compton scattering,''
%  Phys.\ Rev.\ C {\bf 81} (2010) 035202;
%  %%CITATION = ARXIV:0911.0091;%%
%\cite{Mazouz:2007aa}
Mazouz,~M. {\it et al.}  [Jefferson Lab Hall A Collaboration]:
Deeply virtual compton scattering off the neutron.
Phys.\ Rev.\ Lett.\  {\bf 99} 242501 (2007).
%%CITATION = ARXIV:0709.0450;%%

%\cite{Guidal:2010de}
\bibitem{data2}
Guidal,~M:
%``
Constraints on the $\tilde{H}$ Generalized Parton Distribution from 
Deep Virtual Compton Scattering Measured at HERMES.
%,''
  Phys.\ Lett.
\ B {\bf 693}, 17 (2010).
%  [arXiv:1005.4922 [hep-ph]].
  %%CITATION = ARXIV:1005.4922;%% 
  


%\cite{Berger:2001zb}
\bibitem{deu} 
  Berger,~E.~R., Cano,~F., Diehl,~M., and Pire,~B.:
  %``
Generalized parton distributions in the deuteron.
  Phys.\ Rev.\ Lett.\  {\bf 87}, 142302 (2001).
%  [hep-ph/0106192].
  %%CITATION = HEP-PH/0106192;%%
  %72 citations counted in INSPIRE as of 01 Nov 2013
  
\bibitem{3He}
%\cite{Friar:1990vx}
%\bibitem{Friar:1990vx}
Friar,~J.~L. {\it et al.}:
  %``
Neutron polarization in polarized He-3 targets.
  Phys.\ Rev.\ C {\bf 42} 2310 (1990).
  %%CITATION = PHRVA,C42,2310;%%
  %\bibitem{antico}

\bibitem{old}
%\cite{CiofidegliAtti:1993zs}
Ciofi degli Atti,~C. {\it et al.}:
  %``
Nuclear effects in deep inelastic scattering of 
polarized electrons off polarized He-3 and the neutron spin 
structure functions.
%''
Phys.\ Rev.\ C {\bf 48} 968 (1993).
%  [nucl-th/9303016];
  %%CITATION = NUCL-TH/9303016;%%  
  
  
%\cite{Rinaldi:2012pj}    
\bibitem{noi}
Rinaldi,~M. and Scopetta,~S:
  %``
Neutron orbital structure from generalized parton distributions of 3He.
%,''
%  arXiv:1204.0723 [nucl-th].
Phys. Rev. C 85, 062201(R) (2012);
  %%CITATION = ARXIV:1204.0723;%%
%  
%\bibitem{noiarxive}
%\cite{Rinaldi:2012ft}
Rinaldi,~M. and Scopetta,~S:
  %``
Extracting neutron generalized parton distributions from 3He data.
Phys.\ Rev.\ C {\bf 87}, 035208 (2013).
%arXiv:1208.2831 [nucl-th].
  %%CITATION = ARXIV:1208.2831;%%

%\cite{Scopetta:2004kj}
%\bibitem{Scopetta:2004kj}
\bibitem{scopetta}
Scopetta,~S:
  %``
Generalized parton distributions of $^3$He.
  Phys.\ Rev.\ C {\bf 70} 015205 (2004);
Scopetta,~S:   
Conventional nuclear effects on generalized parton distributions of trinucleons.Phys.\ Rev.\ C {\bf 79} 025207 (2009).
%  [nucl-th/0404014];
  %%CITATION = NUCL-TH/0404014;%%
%\cite{Scopetta:2009sn}
%\bibitem{Scopetta:2009sn}
%  S.~Scopetta,
  %``

%  [arXiv:0901.3058 [nucl-th]].
  %%CITATION = ARXIV:0901.3058;%%
  
%\cite{Kievsky:1996gz}
\bibitem{overlap}
Kievsky,~A., Pace,~E., Salm\`e,~G. and Viviani,~M.:
  %``
Neutron electromagnetic form-factors and inclusive scattering of polarized electrons by polarized He-3 and He-3 targets.
  Phys.\ Rev.\ C {\bf 56}, 64 (1997).
%  [nucl-th/9704050].
  %%CITATION = NUCL-TH/9704050;%%

\bibitem{pot} 
%\cite{Wiringa:1994wb}
%\bibitem{Wiringa:1994wb}
Wiringa,~R.~B., Stoks~V.~G.~J. and Schiavilla,~R.:
  %``
An Accurate nucleon-nucleon potential with charge independence breaking.
%,''
Phys.\ Rev.\ C {\bf 51} 38 (1995).
%  [nucl-th/9408016].
  %%CITATION = NUCL-TH/9408016;%%  
  
%\cite{Kievsky:1997bg}  
\bibitem{AV18}
Kievsky,~A., Viviani,~M. and Rosati,~S.:
  %``
Study of bound and scattering states of three nucleon systems.
%''
Nucl.\ Phys.\ A {\bf 577} 511 (1994).
%  [nucl-th/9706067].
  %%CITATION = NUCL-TH/9706067;%%
  
%\cite{Musatov:1999xp}
\bibitem{Rad1}
Musatov,~I.~V. and Radyushkin,~A.~V.:
  %``
Evolution and models for skewed parton distributions.
%''
Phys.\ Rev.\ D {\bf 61} 074027 (2000).
%  [hep-ph/9905376].
  %%CITATION = HEP-PH/9905376;%%
  
%\cite{Marcucci:1998tb}
\bibitem{schiavilla}
Marcucci,~L.~E., Riska,~D.~O. and Schiavilla,~R.:
  %``
Electromagnetic structure of trinucleons.
%''
Phys.\ Rev.\ C {\bf 58}, 3069 (1998).
%  [nucl-th/9805048].
  %%CITATION = NUCL-TH/9805048;%%
  
%\cite{Pace:2013bq}
\bibitem{ema} 
Pace,~E, Salm\`e,~G., Scopetta,~S., Del Dotto,~A. and Rinaldi,~M.:
Neutron Transverse-Momentum Distributions and Polarized $^{3}He$ within Light-Front Hamiltonian Dynamics.
Few Body Syst.\  {\bf 54}, 1079 (2013).
  %%CITATION = ARXIV:1301.5787;%%
  %1 citations counted in INSPIRE as of 03 Nov 2013

%\cite{Belitsky:2000gz}
\bibitem{diet} 
  Belitsky, A.~V., Mueller, D., Niedermeier, L. and Sch\"afer, A.:
Leading twist asymmetries in deeply virtual Compton scattering.
Nucl.\ Phys.\ B {\bf 593}, 289 (2001).


\end{thebibliography}
\end{document}